# 100+ Metrics for Software Startups – A Multi-Vocal Literature Review

Kai-Kristian Kemell[1[0000-0002-0225-4560]], Xiaofeng Wang[2[0000-0001-8424-419X]], Anh Nguyen-Duc[3[0000-0002-7063-9200]], Jason Grendus[4], Tuure Tuunanen[1[0000-0001-7119-1412]], and Pekka Abrahamsson[1[0000-0002-4360-2226]]

[1] University of Jyväskylä, 40014 Jyväskylä, Finland
{kai-kristian.o.kemell, pekka.abrahamsson, tuure.tuunanen}jyu.fi
[2] Free University of Bozen-Bolzano, 39100 Bozen-Bolzano, Italy
xiaofeng.wang@unibz.it
[3] University of Southeast Norway, Norway
angu@usn.no
[4] 3D Ventures Oy, Singapore
jgrendus@gmail.com

**Abstract.** Metrics can be used by businesses to make more objective decisions based on data. Software startups in particular are characterized by the uncertain or even chaotic nature of the contexts in which they operate. Using data in the form of metrics can help software startups to make the right decisions amidst uncertainty and limited resources. However, whereas conventional business metrics and software metrics have been studied in the past, metrics in the specific context of software startup are not widely covered within academic literature. To promote research in this area and to create a starting point for it, we have conducted a multi-vocal literature review focusing on practitioner literature in order to compile a list of metrics used by software startups. Said list is intended to serve as a basis for further research in the area, as the metrics in it are based on suggestions made by practitioners and not empirically verified.

## 1 Introduction

The importance of data in business has greatly increased over the last few decades as acquiring, storing, and using it has become both easier and cheaper in the wake of technological progress. This development was further underlined following the still relatively recent emergence of the big data discourse [47], which encouraged organizations to acquire and store vast amounts of data even if they did not necessarily have any present use for it. Data is now often used by various businesses to support decision-making, even though manager intuition is often in practice still just as important in strategic decision-making [26].

For the purpose of decision-making, data can be used in the form of metrics. Metrics are quantifiable measurements of a phenomenon or object. They are present eve-rywhere in our everyday life from measuring height and weight to measuring speed while driving. Even qualitative data can to some extent be made quantifiable with the right approach: a simple yes or no question can be seen as a Boolean of 1 or 0. In terms of quantifying written statements, techniques such as the Likert scale survey, where users rate qualitative statements on a scale of e.g. 1 to 5 based on how much they agree or disagree with them, have been employed.

Much like larger software companies, software startups can also employ various metrics to measure progress and to aid in decision-making. Given that software startups usually operate under a notable lack of resources and in particularly tumultuous contexts [44], software startups can arguably benefit from the use of metrics. Making the right decisions amidst uncertainty can make all the difference between success and failure. However, based on past survey data[1] from 4700 software startups, most of them in fact did not track metrics or did not use the data gained from tracking them to make decisions. More specifically, 41% of these 4700 software startups felt that it was too early for them to track metrics. Out of the remaining 59% of the responses, some 16% did not track metrics either because they did not have the resources to do or because they did not believe it would benefit them, and 14% tracked them but remarked that the data had no influence on their decision-making.

The majority of software startups end in failure [44]. Arguably, the proper use of the right metrics is something that can help alleviate this situation in part. Metrics can alert a business of approaching disasters and give them time to react before the resulting decrease in revenue really hits them. For example, tracking Daily Active Users (DAU) is a metric that gives near real-time data of how a software is doing. If the number suddenly starts dropping dramatically over the course of a few days, something is likely wrong. Perhaps an update was deployed on the day the initial drop started, and perhaps that update dramatically affected the stability of the software on some devices or operating systems. Nonetheless, in a situation where this hypothetical company was not tracking their DAU, this problem may have only become apparent through a dramatic drop in revenue at the end of the month. However, metrics are typically quite context-dependent; for a very early-stage software startup that is still developing their first product and thus has no users yet, tracking the aforementioned DAU serves no purpose.

Though metrics have been extensively studied in various context across disciplines, metrics specifically in relation to software startups is an emerging area of research. While e.g. classic business metrics such as Net Present Value [38] are certainly applicable to software startups as well, our understanding of what metrics are specifically useful for software startups is presently lacking. To this end, we seek to un-

[1] This was a large-scale survey that ultimately collected 10000+ responses, conducted to explore different aspects of software startups. However, after cleaning the data and filtering it based on whether this particular question about metrics was answered, ~4700 responses remained. As the survey was extensive, most questions were not mandatory, and thus not all responses included answers to all of the questions. Additionally, the numbers are approximations as even after cleaning the data of duplicate or dubious responses (e.g. "name: test.com") no doubt not all of the remaining responses are valid data. Data from the same survey was also used by Wang et al. [48] among others.

derstand what metrics software startups currently use, or are expected to use, based on a multi-vocal literature review focusing primarily on practitioner literature. Through the literature review, we aim to compile an extensive list of potential metrics for software startups, creating fertile ground for further research on metrics in this context. This list is intended to propose potential metrics but offers little insight in which of these metrics *should* be used. Thus, we formulate the research problem of this paper as follows:

**RQ:** What metrics could software startups use to track progress of their business?

The rest of this paper is structured as follows. In the upcoming second section we discuss software startup metrics as an area of research in relation to extant research across disciplines. In the third section we go over the methodology of this study in detail, and in the fourth section we present our results. The implications and limitations of the results are discussed in the fifth and final section that also concludes this paper.

## 2 Software Startups and Metrics

In utilizing metrics, software startups combine various types of metrics. They can utilize conventional business metrics, as well as business metrics more specifically aimed at startups, as well as software-related metrics including website metrics. Across different life cycle stages (e.g. those proposed by Wang et al. [48]), different metrics can be important for software startups. For example, conventional financial metrics are not as relevant for early-stage startups that may still be in the process of acquiring their first customers or that are still calculatingly running a deficit for the time being. A more relevant metric in such a situation could be to simply measure the amount of remaining expendable capital.

Software Engineering (SE), metrics can be split into process metrics and product metrics [49]. Process metrics are metrics related to the process of creating the software, or maintaining it during its operational life, while product metrics are related to the qualities of the product. Product metrics can be seen to include usability-related metrics as well. Process metrics, on the other hand, account for various method-specific or practice-specific metrics such as lean or agile software development metrics [24]. Website-related metrics can also be considered to be a part of SE metrics, however, as websites are ultimately software [49].

In terms of website metrics specifically, basic metrics related to system (website) performance such as site availability or bandwidth [46] have become less relevant in the wake of technological process, particularly following the popularization of cloud technology. It is now virtually a given that a website can handle any ordinary spikes in traffic load with more capacity being allocated as necessary. Indeed, rather than tracking at system-related metrics, the focus from a business point of view has shifted towards understanding the way users interact with it [4]. While assuring system per-formance is no less relevant than before, it is now far easier to achieve website stability with modern computational power.

Organizations aim to comprehensively track the way users use their website in order to better understand them and to optimize it accordingly [4]. Generic metrics for this purpose include tracking visit length per page, tracking what the users click (if anything at all), as well as tracking where the users enter the website from. With large amounts of data becoming increasingly cheap and easy to handle, and with tools for gathering and analyzing such data now being readily available (e.g. Google Analytics), tracking individual users in this fashion has become widespread even among smaller organizations, including software startups. This way of tracking users is not limited to websites. Software companies are equally interested in understanding how the users of their software interact with it in practice in order to improve the software based on the data.

Though software startups occasionally also concern themselves with directly studying usability and User Experience (UX), UX and usability are typically evaluated by actively involving users as participants for a study while either directly observing their use or having the users self-report their experiences through a form. Directly confronting users and potential users in order to better understand their needs can be important and is certainly something software startup practitioners often choose to do as well. However, involving users in order to better understand their needs is something that can be carried out in a similar fashion regardless of whether the organization involved is a software startup or a larger organization. We thus consider them to be out of scope for this literature review as the extant studies in the area are already reasonably applicable to the software startup context as well. This is not to say that further studies on UX and usability testing from the point of view of software startups would not be worth carrying out, however.

As for business metrics, conventional business metrics such as the Net Present Value studied in economic disciplines are also applicable to software startup. However, an early-stage software startup may not yet have a single customer or even a product and thus have no revenue, making many of the more conventional financial metrics less relevant to them especially in their earlier stages. Metrics such as Customer Acquisition Cost, which measures the cost of acquiring a new customer by means of e.g. advertising, can be far more useful for such startups. Similarly, software startups aim for explosive growth and highly scalable business models [44] and thus are also likely to be particularly interested in metrics related to growth over shorter periods of time.

Extant research has extensively studied business metrics, website metrics, and software development related metrics [24] in various contexts. On the other hand, academic research specifically focused on metrics from the point of view of software startups is currently scarce. Software startups are to some extent similar to larger software companies and operate within the same area of the software industry. However, software startups also differ from larger or more mature software organizations in various ways. Thus, while conventional business metrics or software metrics not specifically aimed at software startups are likely to be applicable to software startups, they may not be as important to software startups.

Whereas academic literature on metrics from the point of view of software startups is currently scarce, practitioner literature contains various accounts on software startup metrics. In order to promote discussion and to encourage research in the area, we will review some of the practitioner literature in the area and present the practitioners' views on what metrics software startups should utilize. The details of this multi-vocal literature review are discussed next.

## 3 Methodology

A multi-vocal literature review primarily focusing on practitioner accounts was conducted to collect data for the purpose of formulating a list of preliminary results. As practitioner literature is very heterogeneous in nature, ranging from books to blogs and lacking in common publication platforms such as journals, establishing a fully systematic protocol for reviewing it is challenging due to the vast amount of available data. We nonetheless devised a protocol in order to conduct the review in a semi-systematic fashion. In this case we refer to it as semi-systematic as it consisted of multiple steps, of which the second one was conducted in a systematic fashion.

The literature review consisted of three steps of searching for literature. First, we reviewed popular books written by high-profile practitioner experts (e.g. Eric Ries and Steve Blank) that were relevant from the point of view of metrics. Secondly, we conducted a set of Google searches in order to find less high-profile practitioner literature such as blog posts from various practitioners involved with software startups. Then, using the literature gathered during the first two steps, we finally utilized the snowballing technique to discover more literature discussed in the documents already included for the review.

For the Google searches, we followed a systematic protocol in order to gather higher quality data. The following queries were used for these searches: "software startup metrics", and "startup metrics", "startup metrics list", and "startup what to measure". For each query, the first five pages of results were screened for inclusion. The results were evaluated for inclusion based on the following inclusion criteria:

- The document is not clearly intended as an advertisement for a tool (e.g. a firm writing a blogpost to recommend their own data analytics tool)
- The document presents or discusses specific, actionable metrics (as opposed to non-specific groups of metrics such as sales metrics)
- The document is a textual document and not e.g. a link to a video or a slideshow
- The document is a stand-alone document written under a real name (i.e. not a forum post written under a pseudonym)
- The document is publicly available; not behind a pay-wall or registration
- The document contains metrics that can be employed by most software startups (e.g. not only e-commerce metrics)
- The document is not a duplicate result from another search query

We chose to not limit our inclusions to metrics specifically presented as *software* startup metrics. This choice was made because practitioners seldom speak of software startups. In practitioner literature, startups are typically assumed to be technology companies, or to either be engineering software or be using software to create value for their users. Thus, practitioners seem to think of software startup as a redundant construct when most startups indeed are focused on software. Rather than speaking of software startups, practitioners either simply speak of startups or focus more specifically on e.g. e-commerce startups. On the other hand, SE literature often refers to software startups specifically, and New Technology-Based Firm (NTBF)[2] is a long-standing construct used to refer to startups in business literature. We therefore chose to include documents speaking of startup metrics in general when those metrics were also applicable to software startups, and indeed most such documents not focused solely on financial metrics did discuss user and software metrics.

Finally, in addition to the practitioner literature some general-purpose software engineering metrics were adapted from extant academic literature. For example, some practitioner literature discussed monitoring operational efficiency and time spent on various tasks. We would occasionally adapt such generic, although nonetheless actionable, metrics to be more specific by employing existing research.

In this fashion, we sought to compile an extensive, although by no means comprehensive, list of metrics for software startups based primarily on practitioner literature. These results will be discussed in the following section.

## 4 Results: General-Purpose Software Startup Metrics

Much of the practitioner literature reviewed for this paper consisted of short “n metrics a startup must measure” type lists of five to ten metrics. As a result, there was a considerable amount of overlap. On the other hand, this points to there being some consensus among practitioners as to which metrics are particularly interesting. The most commonly cited metrics were: (1) user churn and user retention metrics, (2) user engagement metrics and metrics measuring user activity, (3) financial metrics focusing on short-term developments and cash burn, and (4) user-focused financial metrics such as User Acquisition Cost.

Churn, in this context, is used to refer to the number of users lost during a time period. The number of total users is important for monetizing any software. However, in the case of freemium software where the software itself is free and revenue is made through ads or in-software purchases, the number of *active* users becomes increasingly important. Such business models are common among software startups and the practitioner literature reflected this in relation to metrics.

In addition to closely measuring the number of users leaving, the activity of the users was regularly cited as an important focus as well. Simply measuring e.g. total users or registered users was considered insufficient. Instead, software startups were regularly urged to focus on measuring at least their Monthly Active Users (MAU) and, more importantly, Daily Active Users (DAU). Other such activity metrics suggested by practitioners were recency, that is, the number of days since the login of a user (i.e. aging / cohort analysis), as well as frequency of logins of the users. Furthermore, while measuring churn, software startups were also encouraged to measure user retention, that is, the number of users coming back to use the software as opposed to permanently leaving.

In addition to simply measuring how often the users used the software, software startups were urged to measure user engagement through various metrics. What exactly constitutes engagement changes based on each software, but in addition to activity, engagement was suggested to be measured by tracking what exactly the users do while using the software. For example, in a digital game, one indicator of user engagement could be the act of actually completing a task (a "quest") in the game as opposed to simply logging into the game, which in and of itself does not verify that a user is in fact doing anything in the game.

Finance-wise, software startups were recommended to focus primarily on user and customer-related metrics alongside more general financial metrics. User or Customer Acquisition Cost (CAC), i.e. the average cost of acquiring a new (paying) user, and User or Customer Life-Time Value (LTV) were the most commonly cited financial metrics. Past the user-focused financial metrics, conventional financial metrics such as revenue and profit margin were commonly discussed, although emphasis was placed especially on metrics indicating more short-term finances such as Month-on-Month growth and Monthly Recurring Revenue. Similarly, (Cash) Burn Rate and metrics related to it (e.g. monthly cash burn) were also commonly recommended for software startup practitioners to utilize. This ties to the fact that software startups are indeed typically lacking in resources, including capital, and are largely reliant on outside funding especially early on in their life cycles [44].

Past these most commonly cited metrics discussed so far, we uncovered a wide variety of metrics intended for software startup use. As our intention was not to study what *should* be measured but what *could* be measured, we chose to include any metrics thought to be relevant enough to be listed in the practitioner literature. To this end, the full list of metrics gathered during the literature review can be found in its entirety in the table below (Table 1), in alphabetical order. A total of 118 metrics were included in the table.

Some of the metrics listed are derivative. E.g. one could simply speak of customer churn in relation to the number of lost customers. However, some writers went into detail about churn-related metrics by discussing monthly churn, net churn and gross churn separately. In these cases, the sub-metrics were listed as well. On the other hand, some metrics were also merged together under more prevalent metrics. For example, "cancellations" [5] was considered related to user churn. Finally, for the purpose of making the table easier to read, only up to three references were included per metric given that e.g. Customer Acquisition Cost was discussed in 18 different references of this paper.

**Table 1.** List of Software Startup Metrics from Practitioner Literature

| Metric and up to 3 Reference(s) | Description |
|---|---|
| Abandonment [12] | Transactions abandoned before completion |
| Acceptance Rate [12] | Avg. no. invites accepted by new users |
| Activation Rate [8, 13, 25] | Number of visitors or users performing a specif- |
| | ic action such as registering or installing |
| Active User Growth Rate [12] | No. new active users in a time period |
| Ad Inventory [12] | Total views of each ad in a time period |
| Ad Rates [12] | Value of each ad. inventory |
| Amplification Rate [25] | No. shares on social media per customer |
| Annual Contract Value [13, 17, 22] | Avg. annualized revenue per customer contract |
| Annual Recurring Revenue [13, 22, 41] | Predictable revenue annually (e.g. subscriptions) |
| Annual Run Rate [13] | Projected annualization of monthly recurring revenue |
| Avg. Revenue per User [13, 15, 25] | Avg. revenue per user over a time period |
| Avg. Revenue per Customer [13, 17, 25] | Avg. revenue per customer over a time period |
| Average Time on Hold [12] | Time user spends on hold when calling support |
| Billings [13] | Current quarter revenue plus deferred revenue from previous quarter |
| Bounce Rate [8, 40] | Percentage of visitors leaving website quickly |
| Breakeven Analysis [3] | Analysis to determine the point where revenue covers the costs of receiving it |
| Burn Rate [8, 15, 18] | Rate at which available capital is used |
| Campaign Contribution [12] | Added revenue from an ad campaign |
| Capital Raised to Date [23] | Amount of investment capital raised in total |
| Cash Flow Forecast [3] | Forecast of financial liquidity in a period of time |
| Cash on Hand [19] | Available capital |
| Churn Rate [1, 15, 17] | Lost users or customers over a time period |
| Click-Through Rate [12] | Visitors that clicked a specific website link |
| Committed Weekly Recurring Gross Profit [45] | Percentage increase in profits weekly committed recurring profit |
| Compounded Monthly Growth Rate [13] | Avg. % growth per month since inception, or another start point for measuring. |
| Content Creation [12] | No. visitors that interact with website content |
| Conversion Rate [1, 8, 17] | No. visitors that become users or customers, or no. users that become customers. |
| Cost of Goods Sold [23] | Cost of products or services sold (e.g. hosting) |
| Customer Acquisition Cost [3, 7, 8] | Average cost of acquiring a paying user. |
| Customer Acquisition cost to life-time value ratio [11, 30] | Customer Acquisition Cost vs. Customer Life-time Value |
| Customer Concentration [13, 31] | Revenue from largest customer vs. total revenue |
| Customer Count [39] | Total number of customers (paying users) |
| Customer Retention Cost [25] | Amount of spending on customer retention |
| Daily Active Users [9, 11, 13] | No. users who use the software daily |
| Daily Active Users to Monthly Active Users ratio [25] | A more detailed measure of user activity |
| Deferred Revenue [13] | Revenue received in advance of earning it |
| Development Time [18, 39] | Time it takes to implement a new feature |
| Direct Traffic [13] | Traffic coming in directly |
| Downloads or Installs [22] | Total amount of downloads or installs |
| E-mail Conversion Rate [34] | Number of recipients that e.g. became users |
| E-mail Open Rate [34] | No. mailing list members that open an email |
| Facebook Likes [5] | Number of likes on firm Facebook page |
| Fixed vs. Variable Costs [3] | A measure of total spending split by source. |
| Frequency of Logins [17] | Average frequency of user logins |
| Frequency of Visits [25] | Average frequency of visits to e.g. website |
| Gross (Cash) Burn [13] | Monthly expenses and any other outlays |
| Gross Churn Rate [13, 37] | Total users lost |
| Gross Margin [7, 13, 15] | Total revenue compared to cost of goods sold |
| Gross Profit [13, 17, 22] | Total revenue minus cost of goods sold |
| Innovation Metabolism [14] | Number of build-measure-learn cycles |
| Intent to Use [28, 34] | Data indicating that a new user is about to start using the software. E.g. imported custom data |
| Invitation Rate [12] | Avg. no. invites sent per existing user |
| Launch Rate [12] | No. downloaders that launched the software |
| Leads [29] | An estimate of prospective customers. |
| Lead-to-Customer rate [29] | Number of leads converted into customers |
| Life-time Value [3, 7, 8] | The average total revenue a customer generates |
| Likes per Post [34] | Likes per social media post |
| Load Time [9] | Time it takes for software to start or respond to user commands |
| Market Share [50] | |
| Market Value [50] | |
| Monthly Active Users [8, 9, 11] | No. users who use the software monthly |
| Monthly Cash Burn Rate [13, 30] | |
| Monthly Churn Rate [13] | Lost users or customers per in a month |
| Monthly Recurring Revenue [10, 11, 13] | Monthly predictable revenue (e.g. subscriptions) |
| Month-on-Month Growth [10, 13, 17] | Average of monthly growth rates |
| Net Adds [12] | Total new customers vs. cancellations |
| Net (Cash) Burn Rate [13] | Gross cash burn vs. revenue in a period of time |
| Net Churn [13] | New users gained vs. users lost |
| Net Promoter Score [9, 13, 17] | How likely users are to recommend product |
| Network Effects [13] | Effect of one user on the value experienced by other users (e.g. Metcalfe's Law) |
| New Visitors [17] | Number of new visitors |
| Number of Logins [5, 13] | Logins per user over a period of time |
| Number of Transactions [39] | Number of transactions made in a time period |
| Office Morale [5] | How motivated the team is |
| Operation Efficiency [15, 18] | Comparison of firm expenses by source |
| Organic Traffic [13] | Unpaid traffic from e.g. Google search results |
| Payback Time [25] | Time to recoup from an expense via revenue |
| Payment failures [45] | Number of failed transactions from users |
| Platform Risk [13] | Dependence on a specific platform or channel |
| Profit Margin [17, 25, 30] | Revenue minus cost divided by revenue for a product. Different ways to measure for e.g. Software-as-a-Service companies. |
| Prospects [12] | Number of users that might become customers |
| Purchases [12] | No. purchases made by a user in a time period |
| Recency [21] | Days since last visit of user |
| Referrals from current users [8, 27, 31] | How often current users refer new users |
| Referral rate [1] | Volume of referred users or purchases |
| Registered Users [17] | Total number of registered users |
| Repurchase Rate [23] | No. customers that made a purchase during the previous and current period of time |
| Retention Rate [1, 7, 8] | Percentage of users or customers still using the service after a period of time |
| Retention by Cohort [13] | % of original user base still using the software or conducting transactions in it |
| Return on Advertisement Spending [7] | Profits divided by advertisement spending |
| Revenue [5, 17, 22] | Total Revenue |
| Revenue Growth Rate [41, 43] | |
| Revenue Run Rate [11, 15] | |
| Reviews Considered Helpful [12] | Number of reviews considered helpful |
| Reviews Written [12] | Number of reviews written |
| Sell-through rate [13] | No. units sold in a time period in relation to the no. items in inventory at its beginning |
| Session Interval [17] | Average time between software use sessions |
| Session Length [17] | Length of average software use session |
| Social Media Reach [34] | Post reach within e.g. Twitter or Facebook |
| Sources of Traffic [17, 27, 31] | Source and volume of user traffic per source |
| Stability [9] | Frequency of crashes in software use |
| Time to Customer Breakeven [12, 30] | Time it takes to recoup from Customer Acquisition Cost |
| Time to First Purchase [12] | Avg. time users take to become customers |
| Top Keywords Driving Traffic to You [12] | Search terms used by visitors to find your site |
| Top Search Terms [12] | Both those that lead to revenue, and those that don't have any results. |
| Total Ad Clicks [12] | Number of advertisements clicked by visitors |
| Total Addressable Market [13, 17, 50] | Total hypothetical market size |
| Total Contract Value [13, 17, 22] | Value of one-time and recurring charges |
| Total Number of Customers [8, 32] | |
| Total Number of Users [5, 50] | Based on e.g. registered user accounts |
| Traffic [1, 5, 18] | Total number of website visits (non-unique) |
| Traffic-to-Leads [1] | Total traffic in relation to potential customers |
| Uptime [40] | Percentage of time software or website is available and operational |
| User Acquisition Rate [5, 9] | Total new non-paying users in a time period |
| User Demographics [5, 9] | Avg. age, gender distribution, location etc. |
| User Engagement [9, 17, 28] | Measured through e.g. login frequency. Definition depends on context. |
| Unique Visitors [11] | Unique website visitors during a time period |
| Viral Coefficient [11, 13, 32] | No. new customers each existing one converts |

While the metrics listed above (Table 1) are applicable to most software startups, all metrics are ultimately context-specific to some extent and thus more useful for some software startups than others. Furthermore, metrics specifically targeted at smaller sub-sets of software startups can be more insightful to firms belonging to that sub-set than general-purpose business metrics for software startups. An e-commerce company will likely be focusing specifically on metrics related to their online store or platform, even though more universal software metrics such as Daily Active Users can supplement that data.

Furthermore, in terms of software engineering related metrics, practice-specific and method-specific metrics can be highly relevant to an organization. That is, if the work is not done ad hoc as it occasionally is in software startups [35]. Various agile methods and practices have their own metrics either built into the method (e.g. sprint duration in Scrum) or metrics for them have been suggested by extant research (e.g. [24]). Though such method-specific metrics can be applicable to any software startup choosing to employ a particular method, they are arguably not universally applicable to software startups. Methods and practices used to engineer software are highly diverse, with practitioners often choosing to use in-house methods created by tailoring existing methods and practices [16]. This is also the case for software startups [35]. Indeed, few method-independent SE metrics were discussed in the literature.

## 5 Discussion and Conclusions

In this paper, we have presented an unverified list of software startup metrics primarily based on practitioner literature (Table 1). Though we have provided an extensive list of various metrics for software startup practitioners, we have offered little verification for any of the listed metrics. The list can offer ideas for what to measure but cannot verify what effect tracking any of these metrics may have for a software startup. We can also not offer any recommendations on which metrics to use to achieve different goals. Furthermore, though the list is extensive, it is not comprehensive: many other metrics, especially more context-specific ones, can be conceived. Additionally, various conventional SE metrics and financial metrics not included in the list can likely be applied to software startups even though they were not present in the literature reviewed.

Another issue with the data is that many of the practitioner accounts dealing with software startup metrics come from the point of view of third parties. I.e. rather than being written by software startup practitioners for software startup practitioners, many of the writers are investors, startup advisors, and other external affiliates. Thus, many of the metrics discussed in the practitioner accounts reviewed for this paper were metrics (potential) investors typically wish to see when considering investing in a software startup. On the other hand, some of the practitioner accounts also discussed metrics mainly intended for internal organizational use in software startups such as operational effectiveness.

Furthermore, data and metrics are powerful tools but need to be utilized in a fitting fashion to be useful. It is important to measure relevant phenomena and to use the data to make decisions in a context-dependent fashion. More universally applicable metrics such as the ones presented in this paper can offer a useful starting point for practitioner organizations. However, more context-specific metrics such as e-commerce startup metrics can offer more valuable insights inside that context. Furthermore, every company can devise metrics unique to that company specifically that may offer even better insights into their business specifically. For example, a software startup whose main product is an online game may use metrics related to in-game data from that particular online game in order to improve the product.

Nonetheless, despite its limitations, the list of metrics presented in this paper is both a part of on-going research as well as a research proposal. Those interested in software startups and their use of metrics can make use of this list in further studies in that area. Further research on the topic could seek to study some individual metrics or groups of metrics in empirical settings, or to categorize the metrics to better suit certain contexts such as the aforementioned e-commerce domain while also adding more context-specific metrics related to that area.

On the other hand, practitioners affiliated with software startups may utilize the list to potentially gain new insights into what metrics software startups could measure. We urge any interested practitioners to view the list through the lens of their particular business and to use their own judgment on which metrics could be potentially relevant for their business. While there exists some consensus on what is important to measure in software startups in the practitioner literature reviewed for this study, we can currently offer no empirical validation in favor of any of them.

To summarize, we conducted a multi-vocal literature review primarily focused on practitioner literature. We combined an extensive list of software startup metrics (Ta-ble 1 in section 4) that software startups could measure. Based on the literature, practitioners generally recommend that software startups focusing on measuring:

- User retention and user churn
- Active users and user engagement
- Short-term focused financial metrics such as month-on-month growth and cash burn rate
- User-focused financial metrics such as User Acquisition Cost

While there was a large amount of variety in the metrics discussed in the practitioner literature, these were the most prevalent metrics among the literature reviewed. However, ultimately every business is unique and needs to establish separately which metrics are relevant for that particular business. Similarly, different metrics serve different purposes. Financial metrics may serve to indicate that something is wrong with a software but will likely not help in understanding what that might be.